# 1. Executive Summary – An X-ray Polarimeter for Constellation-X

K. Jahoda, K. Black, P. Deines-Jones, J.E. Hill, T. Kallman, T. Strohmayer, J.H. Swank

Point of contact: keith.m.jahoda@nasa.gov

Polarimetry remains a largely unexploited observational technique in X-ray astronomy which could provide insight in the study of the strong gravity and magnetic fields at the core of the Constellation-X observational program.

Adding a polarization capability to the Constellation-X instrumentation would be immensely powerful. It would make Constellation the first space observatory to simultaneously measure *all* astrophysically relevant parameters of source X-ray photons; their position (imaging), energy (spectroscopy), arrival time (timing), and polarization. This would provide unprecedented leverage for modeling X-ray emission from a whole range of source types, and could break frustrating degeneracies in cases where imaging timing, and spectroscopy are not definitive. Magnetic fields and scattering are ubiquitous processes in astrophysical sources and they naturally lead to polarized X-ray fluxes. In addition, the strong gravity of black holes imposes unique changes of direction of polarized flux from a disk, which should apply to stellar black holes in "high-soft" states and to the reflection and fluorescence from the disk in AGN in their "low-hard" states. Scattering off electrons in slab-like geometries has signatures which could resolve current ambiguities about the geometry and dynamics of these coronae. The polarization from synchrotron emission and scattering could resolve our ambiguities about Blazar X-ray emission regions and allow us to understand the connection of highly collimated jets to the black hole. These objects are so variable, that it would be extremely valuable to have the polarization measurements at the same time as spectroscopy and timing information. Neutron stars are a bridge from matter as we know it in the lab to black holes. Polarization provides an important probe of their diverse magnetic fields, ranging from $10^8$ to $5 \times 10^{14}$ G, and their compact structure imparts polarization signatures that could tell us whether or not their centers contain not only macroscopic nuclear matter, but even quark matter, as it adjusts to being forced to approach collapse to a black hole.

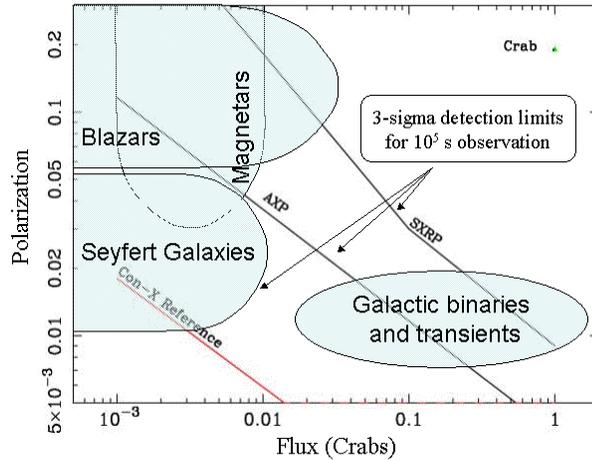

Figure 1: The fractional polarization detection threshold for observations of 100 ksec. The Spectrum XΓ X-ray Polarimeter (SXRP) was built but not flown; the Advanced X-ray Polarimeter (AXP) was a proposed Small Explorer using gaseous photoelectric track imaging polarimeters. The Constellation-X improvements (red) are due to collecting area *and* increased polarimeter efficiency. An $E^{-2}$ spectrum is assumed throughout. The Constellation-X sensitivity assumes that polarimeters are inserted behind two of the four mirror modules.

Astrophysical polarimetry requires sensitive well-calibrated instruments. Many exciting objects are extra-galactic (i.e. faint) and may have small polarization. Recent advances in efficiency and bandpass make it attractive to consider a polarimetry Science Enhancement Package (SEP) for the Constellation-X mission. Figure 1 presents the 3σ sensitivity limits for an SEP which is orders of magnitude more sensitive than SXRP, the most sensitive astrophysical polarimeter ever built (Kaaret 2004), which was however never flown. The sensitivity limits in this figure are based on recent laboratory measurements (Figure 2). In addition to sensitive polarimetry, the SEP provides simultaneous timing from the polarimetry detector and spectroscopy from the primary Constellation-X focal plane instrument. The instrument can follow the change in polarization through an AGN flare or allow folding the data on a Quasi-Periodic Oscillation to study phase effects.

We review the diagnostic power of polarimetry (sec. 2), recent instrumentation advances (sect. 3), the proposed SEP (sec. 4), and its technological readiness (sec. 5).

## *2. Scientific Benefit of Polarimetry*

This section provides an overview of polarimetry's expected contributions to the study of strong gravity via observations of black holes and strong magnetic fields via observations of neutron stars.

**Black Holes**

**Overview of the black hole environment**: A prime objective of Constellation-X is the study of "strong gravity," which is found in the neighborhoods of the event horizons of black holes. What is the physics here really like? The strong gravity controls the propagation of photons and rotates the polarization of a photon as it bends its direction. The Fe-lines from sources which are super-massive black holes can be explained in terms of gravitational red-shifts and Doppler shifts of disk material at the innermost stable orbit of Kerr black holes (e.g. Brenneman & Reynolds 2006; Minniutti et al. 2003). Tests of this picture, as well as of our understanding of General Relativity (GR), would come from measurements of X-ray polarization as a function of energy.

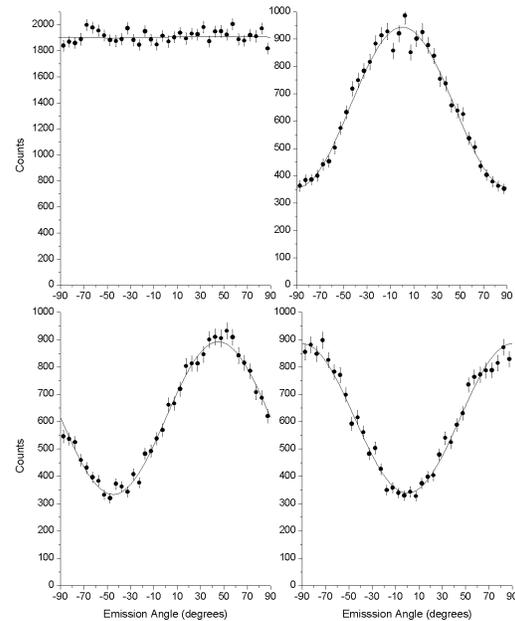

Figure 2: The modulation response of Goddard's demonstration polarimeter to unpolarized 5.9 keV Fe-55 X-rays (upper left) and 100% polarized X-rays with the detector in three orientations with respect to the beam. The measured polarization of the Fe-55 X-rays is 0.49±0.54%. The other three measurements are consistent with 45±1.1% modulation. The detector was filled with 460 Torr of a 50/50 Ne/dimethyl ether mixture.

The starting point for models of accretion onto black holes is an optically thick accretion disk. Its inner radius can be as close as the Innermost Stable Circular Orbit (ISCO), which depends on the black hole's mass and angular momentum. But observations and theory suggest that this picture is too simple. Radiation and magnetic pressure will act to drive mass loss from the black hole either in a broadly distributed outflow or in collimated jets (See review by Konigl 2006). Emission from the disk, the outflows, and from the plunging region between the ISCO and the event horizon (Beckwith, Hawley, and Krolik 2006) must all be considered. Polarization measurements of the fluoresced iron-lines and the continuum would help determine the contributions of the possible processes. The energy dependence of both the degree and angle of linear polarization would be especially useful.

**Inverse-Compton radiation from coronal clouds or outflows:** The spectra and the variability properties of Seyfert galaxies and the stellar black hole Cygnus X-1 in its "low-hard" state are similar, if scaling with the mass is considered (e.g. Uttley and McHardy 2001). The power-law emission from the stellar black holes in this state has long been modeled as thermal Comptonization of photons from a disk up-scattered by high temperature electrons in a corona around the black hole. Evidence of the disk in soft X-ray emission has usually been interpreted as implying the disk is truncated outside of the ISCO, at 100-1000 $GM/c^2$ (with G the gravitational constant, M the mass of the black hole, and c the velocity of light). AGN spectra have been understood to have a similar inverse Compton origin. The low energy wing of the fluoresced iron-line, which extends to ~4 keV, is consistent with the AGN disks extending down to the ISCO. The corona of hot (30-200 keV) electrons could have a spherical geometry or a slab geometry, sandwiching the black hole's equatorial plane and any inward extension of a disk. Many AGN are associated with compact radio sources. Radio emission is also observed to be correlated with the X-ray continuum emission from galactic transient sources (e.g. Gallo et al. 2003). It has been proposed that in both cases the corona is actually the base of a compact jet.

Models of the polarization for both spherical and slab coronae and for the outflows been computed (Matt, Fabian, and Ross 1993; Poutanen and Svensson 1996; Bao et al. 1997; Beloborodov 1998) with polarizations ranging from a few percent to 20%, depending on the optical depth of the corona and the inclination angle of a slab.

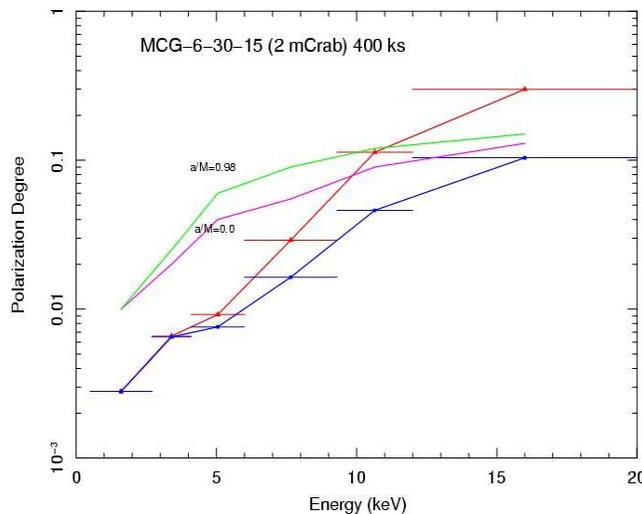

Figure 3. Polarization of an AGN with reflection from a disk source which is on axis a distance of $2GM/c^2$ above the plane. The bars show the polarization detection limits for the energy range of the single Ne/DME detector (red) and the two detector, Ne/dimethyl ether and Ar, combination (blue). The green and magenta curves are for illumination of the disk by an unpolarized coronal source (See text).

**Self-lensing:** When the emission region is very near the black hole, such as a disk in the "high-soft" or "thermally dominant" state of stellar black holes (Remillard and McClintock 2006), the photon trajectories are bent by the gravity of the black hole and frame-dragging of a rotating black hole can cause large rotations in the polarization (Connors, Piran and Stark 1980). Recent work on the continua spectra from disks for stellar black holes implies the angular momentum of GRS 1915+105 is near maximal (McClintock et al. 2006). The inclination angle is about 60 degrees. The polarization would be a few percent to 15% by Newtonian physics or GR, but with different behavior of the polarization as a function of energy. For GR the polarization would rotate from 15 to 30 degrees going from 3 keV to 10 keV, while for Newtonian gravity there would be no rotation. The magnitude should decline from about 2.5 % to 1 % with increasing energy. At higher energy the observer sees a mixture of photons issuing from in front of the black hole mixed with those suffering rotation on their journey from behind the black hole. Bright stellar black hole transients occur at the rate of about 2 per year and would be found by soft or hard x-ray monitors. Recurrences with time scales of 1-50 years of about 20 sources, would make it feasible to monitor them in radio or x-ray bands, a project that might be undertaken with Constellation-X for spectroscopic goals in any case, in order to study quiescent evolution of Advection Dominated Accretion Flows (ADAFs).

While there have been indications that for the galactic black holes in the low state, the disk does not always extend to the ISCO, it is thought to do so for the temperature conditions of the much larger super-massive black holes. In this case the continuum might emanate from above the disk and from a relatively symmetric distribution. The fluoresced line and the reflection spectrum would come from the disk and be subject to the lensing effect. The polarization, like the red-shift and Doppler smearing, would depend on the height above the disk of the non-thermal continuum and the angle of inclination (Matt 1993; Matt, Fabian & Ross 1993; Dovciak, Karas & Matt 2004). The predicted polarization again ranges from tenths to tens of percent and for significant ranges of conditions, the amount and angle depends strongly on energy in the 3-15 keV range. The polarization must in this case be considered as the net of the polarization of the non-thermal continuum (discussed above) seen directly and the second hand fluorescence and reflection.

It is expected that flares are associated with hot spots spiraling into the black hole. Polarization magnitude and direction changes will accompany each circuit (Bao et al. 1997). The polariaztion may exhibit quasi-periodic oscillations which would be directly observable in AGN flares; for stellar mass black holes the phase dependence of the polarization could be studied by folding at a QPO frequency derived from the same data.

**Jet-dominated AGN (Blazars):** Blazar emission is interpreted to be due to relativistic and highly collimated jets, viewed along the axis or obliquely to it. The radius from the black hole at which the jet emerges, the way the magnetic field may thread the black hole's event horizon, and the temperature of the matter in the jet influence the polarization and are crucial to understanding the effect of a black hole on matter near it (Begelman and Sikora 1987; Poutanen 1994; Celotti and Matt 1994; Wolfe and Melia 2006). In these cases polarizations of 40% are typical and even as high as 70%, depending on parameters that include the viewing angle, the uniformity of the magnetic field, and the electron energy distribution.

**Summary:** What we know about sources of strong gravity leads us to believe the polarization should be stronger than 10% for some sources, 0.1-1 % for others and that the determinations will contribute significantly to the strong gravity measurements. Estimated instrument detection thresholds (3 σ) in the energy ranges 2-6 keV and 6-12 keV are better than 3% for a 1 mCrab Crab-like power-law in 50 ks of observations (1 day at an efficiency of 58%). In 3 days, for a 3 mCrab source the sensitivity would reach 1%. There are tens of Seyferts, quasars, and Blazers that would fall in that category. Very significant polarimetric measurements would be made. They could verify (by rotation of the linear polarization with energy) that black holes are at the centers of accreting black hole candidates in the high-soft state and at the center of both stellar and super-massive black holes in low-hard states. They could determine (measuring an energy dependent polarization magnitude) whether power-law continua in the low-hard (and intermediate) states were due to spherical or slab coronae or the bases of relativistic outflows. They could help determine characteristics of the collimated outflows in Blazars, from the viewing angles of the jets to the position of the radio emitting electrons. The stellar black holes have changes of state and the AGN have flares. The polarization measurements would be sensitive enough to see 5 minute changes in bright black holes (> 300 mCrab) and day changes in bright Blazars (50 mCrab).

### Neutron Stars

**Overview.** Neutron stars provide unique laboratories for the study of matter and radiation under the most extreme conditions in the universe. They contain the most dense matter and strongest magnetic fields known. The properties of matter at such high densities are still largely a mystery. For example, the cores of neutron stars may contain rare states of quark matter not found anywhere else in the universe. Our ignorance reflects an inability to uniquely solve from first principles the structure of matter within the theory of strong interactions, Quantum Chromodynamics. One of the few ways to obtain additional constraints on the theory is to accurately determine masses and radii of neutron stars. A major goal of Constellation-X observations is to constrain the properties of superdense neutron star matter and search for evidence of new or exotic states of matter. As we outline below a polarization capability coupled with the large collecting area of Constellation-X would provide a powerful new probe of neutron stars.

Efforts to measure neutron star properties depend largely on detecting radiation directly from their surfaces. Observations of gravitationally redshifted spectral lines, for example, can place accurate constraints on the mass to radius ratio, $GM/c^2R$ (also called the compactness), of neutron stars (Cottam, Paerels and Mendez 2002; Chang et al. 2006). Detection of the Doppler effect associated with rotational motion of the neutron star, as can be done by measuring widths of surface spectral lines, can provide complementary constraints on neutron star radii (Villareal and Strohmayer 2004). Accurate timing of photons from neutron star surfaces, in the context of measuring the amplitude and shape of pulsations at the spin period of the neutron star, can also provide constraints on neutron star structure (Nath, Strohmayer and Swank 2002; Bhattacharyya et al. 2005).

**Emission from the Neutron Star Surface.** A key physical attribute of the photon emission from neutron star surfaces, not yet been exploited to constrain neutron star structure, is polarization, particularly in the context of strongly magnetized neutron stars with surface fields greater than about $10^{12}$ G. In the strong magnetic fields of these objects the scattering and propagation of photons is intimately coupled to polarization. The primary reason for this is that it is much easier to scatter an electron in a direction along the magnetic field than perpendicular to it. For example, in the magnetized plasma that characterizes a neutron star atmosphere, an X-ray photon can propagate in two normal modes: the ordinary mode (O-mode) polarized parallel to the field and the extraordinary mode (X-mode) polarized perpendicular to the field. For photon energies much less than the cyclotron energy, typically about 12 keV for characteristic field strengths, the mean free path of X-mode photons is much longer than that of the O-mode. Hence the X-mode photons stream out from hotter and deeper layers of the atmosphere, and the emergent radiation is highly polarized, with the local direction of polarizion reflecting the local magnetic field direction (Pavlov and Zavlin 2000).

While the emission from a localized patch on the neutron star surface is likely highly polarized, the flux detected by a distant observer represents a sum over all surface elements visible to the observer. Now, the fraction of a neutron star's surface which is visible at infinity is strongly influenced by the gravitational deflection of light rays. The strength of this light bending is proportional to $GM/c^2 R$, with more compact stars having more visible surface area. Thus the polarization fraction measured by a distant observer encodes information about the compactness of the neutron star. The effect is such that more compact stars will have lower overall polarization fractions. The spectrum of polarization, that is, how the polarization fraction changes with photon energy, is sensitive to the strength of the magnetic field as well as the inclination of the magnetic axis. If the star rotates, then measurements of the polarization fraction and position angle with rotational phase can be used to constrain the inclinations of the rotation and magnetic axes (see for example, Pavlov and Zavlin 2000). Models indicate that polarization fractions in the 10 - 35% range can be expected from thermally emitting neutron star surfaces. Higher values are expected for phase resolved measurements

**Vacuum Polarization.** While the above effects are always present, they assume that nothing else influences photon propagation or polarization. However, an additional effect of Quantum electrodynamics (QED) predicts that in a strong magnetic field the vacuum itself becomes birefringent. This prediction of QED has never been tested, but it may reveal itself in the radiation from magnetized neutron stars. This vacuum polarization modifies the dielectric properties of the vacuum and hence the polarization of photon modes, thereby influencing the opacities of photons propagating from the star. Several observational effects are possible. First, the average polarization fraction of neutron star surface emission is predicted to be larger than without the QED effect because vacuum polarization decouples the polarization modes of photons leaving the surface (Heyl, Shaviv, and Lloyd 2003). Second, for field strengths common in many neutron stars vacuum polarization gives rise to a unique energy dependent polarization signature (Lai and Ho 2003). That is, the plane of linear polarization of photons with energies less than about 1 keV is perpendicular to that of photons with energies greater than 4 keV. While the baseline detector we describe is not sensitive below 1 keV, a transition may still be inferred, especially for the strongest magnetic fields. Although many predictions of QED have been tested in terrestrial labs the prediction of vacuum polarization in strong fields has not, and X-ray polarization measurements of magnetized neutron stars may be the only way to do so.

**Observational Prospects.** The classes of neutron stars for which X-ray polarization measurements with Constellation-X would likely be most informative are the thermally emitting isolated neutron stars, and the "magnetar" candidates, the Soft Gamma-ray Repeaters (SGRs) and Anomalous X-ray Pulsars (AXPs). For example the persistent emission from AXPs and SGRs have a spectrum which can heuristically be modelled with a 0.4 - 0.6 keV black body, and a steep power-law with index between 2 and 4 (Woods and Thompson 2006). These spectra peak in the X-ray band in which our polarimeter's sensitivity is well matched to the Soft X-ray Telescopes (SXT) of Constellation-X (2 - 10 keV). Fluxes of these sources span the range from a few tenths to a few mCrab, and our estimates suggest that polarization as low as a few percent can easily be reached in 100 ksec for many of these objects. Because their expected intrinsic polarization is high (perhaps many tens of percent), it is likely that both sensitive phase and energy resolved polarization measurements will be possible (most likely in the 2 - 7 keV band), and almost certainly for the brightest objects. These measurements could provide snapshot "views" of the magnetic field geometry, and perhaps show how it evolves over time. For example, measurements of the polarization properties before and after an AXP or SGR outburst could elucidate details of the magnetic field evolution known to occur in these objects. While the intrinsic surface emission is likely highly polarized, some of these objects may require more complex modelling to fully extract all of the physics because strong and dense magnetospheric currents may be present which can influence the spectrum and polarization via scattering of emergent photons. There has been substantial progress in modeling such effects recently (see Lyutikov and Gavriil 2006; Fernandez and Thompson 2006).

While the thermally emitting, isolated neutron stars are also expected to have substantial intrinsic polarization, these objects have cooler (~40 - 80 eV) thermal spectra which are less well matched to the nominal > 2 keV polarization sensitivity of our instrument. Nevertheless, a few of the hotter objects, such as PSR J0538+2817, could provide important constraints on the average polarization expected for thermally emitting neutron stars.

## *3. TPC Polarimeter Concept*

Photoelectric X-ray polarimetry with finely spaced, pixelized gas detectors has matured into a powerful and practical technique for astronomical observations. In 2003, Black et al. demonstrated the first gas pixel polarimeter suitable for use at the focus of a conical foil mirror. Based on that technology we proposed the Advanced X-ray Polarimeter (AXP) to NASA's Small Explorer program. AXP received the program's highest science rating, and was awarded technology development funding to bring gas pixel polarimeters to greater flight readiness. As a result of those efforts and others, gas pixel polarimeters have now reached a mature level of development (Bellazzini et al. 2006; Hill et al. 2006).

Nevertheless, gas pixel polarimeters are fundamentally limited to quantum efficiencies of ~10%. We recently demonstrated the Time Projection Chamber (TPC) as a photoelectric polarimeter without this sensitivity limit. Like the gas pixel detector, this new polarimeter forms images of photoelectron tracks to extract the polarization information. Our small demonstration polarimeter is comparable to the best pixel polarimeters in intrinsic polarization. The TPC technique can be extended to near unit quantum efficiency at energies of astrophysical interest without loss of sensitivity.

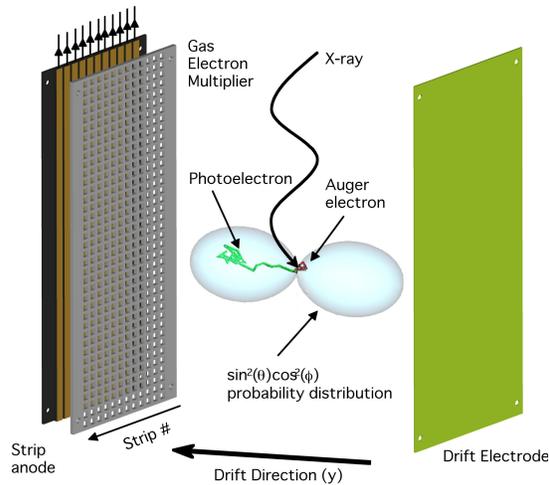
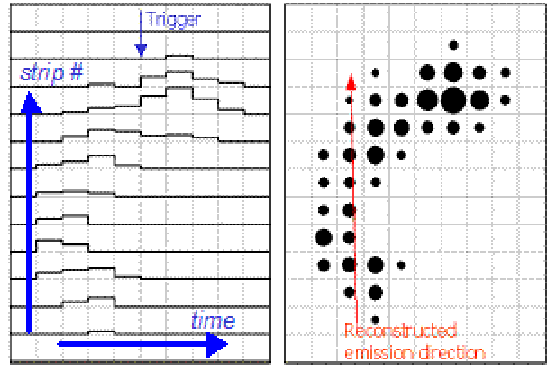

Figure 4a: The time projection polarimeter uses a simple strip readout, left, to form pixelized images of photoelectron tracks.

Figure 4b. The digitized wave forms are interpreted as the density of charge along the photoelectron track. In this image, the horizontal axis is time and the vertical axis is readout strip. The size of the dot is proportional to charge.

The TPC polarimeter instruments a volume of gas to image photoelectron tracks produced by X-rays. An X-ray absorbed by the K-shell of an atom emits a photoelectron in a direction that tends to be perpendicular to the X-ray's incident direction and parallel to the X-ray's electric field vector. A uniform electric field is applied in the gas-filled volume to drift ionization electrons formed along the photoelectron track into a Gas Electron Multiplier (GEM), which amplifies the ionization (Figure 4a). The multiplied charge is deposited on a readout anode connected to charge-sensitive electronics, and the resulting signal is interpreted as an image (Figure 4b). For each strip, the electronics record the waveforms of the charge signals before and after a trigger signal from the GEM. A track image is formed in the plane perpendicular to the GEM by binning the charge pulse trains into pixels whose coordinates are defined by strip location in one dimension and arrival time multiplied by the drift velocity in the orthogonal dimension. We derive the magnitude and direction of a source's polarization by measuring an ensemble of photoelectron directions and constructing a histogram of emission angles (Figure 2). The polarization is related to the modulation $(N_{max}-N_{min})/(N_{max}+N_{min})$ of this histogram, and the polarization direction is the angle corresponding to $N_{max}$.

The key advantage of the TPC over the pixel polarimeter is that its quantum efficiency is independent of its modulation response. The depth of the interaction volume determines the quantum efficiency, while the drift distance determines the amount of electron diffusion, which limits the track image resolution. Behind a grazing incidence X-ray (i.e. high f-number) mirror a deep detector is possible without requiring a large drift distance and image resolution is independent of the interaction depth. In a pixel polarimeter, X-rays enter through the drift electrode; higher efficiency requires increasing the drift distance which reduces track image quality and modulation response.

The TPC also has a simple readout with high rate capability. The TPC readout anode can be fabricated as a ceramic printed circuit board, and the electronics can be implemented with discrete commercial-off-the-shelf electronics. The TPC signals are continuously digitized and pipelined. This scheme incurs a deadtime of less than 1 microsecond per event.

In exchange for higher sensitivity and a simpler readout, the TPC trades focal plane imaging. Since detection occurs when the charge arrives at the GEM, not when the X-ray interacts, the distance between the interaction point and the GEM is unknown.

**Demonstration Polarimeter.** The demonstration TPC polarimeter which produced the data in Figure 2 consists of a GEM with 75 micron diameter holes on a 150 micron, hexagonal spacing. The active area is 3 cm x 1.3 cm. Readout strips, 3 cm long on a 132 micron pitch (Figure 6), are placed 0.5 mm behind the GEM. This readout plane is a standard printed circuit board. The drift electrode is 2 cm away from the GEM.

There are 96 readout strips that are grouped into four sets of 24, giving a total active area of 30 mm × 12.7 mm (depth x width). Every 24$^{th}$ strip is tied together and connected to an analog-to-digital converter through a charge-sensitive amplifier. In this way, the entire active area can be read out with only 24 electronics channels. As long as the valid electron tracks cross fewer than 24 strips, the tracks can be reconstructed without confusion. This scheme can be expanded in both length (detector depth) and width (number of sets of strips) without increasing complexity or power consumption significantly. The strips are read out with commercial preamplifiers and A/D converters.

| | |
|---|---|
| Number of polarimeters | 2 |
| Active volume dimensions | 2.5 x 2.5 x 15 cm$^3$ |
| Fill gas | 50% Ne, 50% dimethyl ether |
| Pressure | 1 atm |
| readout pitch | 75 microns |
| GEM pitch | 150 microns |
| QE @ 6 keV | 43% |
| modulation @ 6keV | >0.5 |

Table 1: Key parameters of the reference polarimetry SEP

The demonstration polarimeter has a polarization response (Figure 2) comparable to the best pixel polarimeters, and has several times higher quantum efficiency. We are currently fabricating a prototype polarimeter with 75 micron readout pitch that will achieve similar modulation at 1 atm.

## *4. Proposed Science Enhancement Package*

The polarimetry SEP consists of two TPC polarimeters and a mechanism to move the instruments onto and off of the optical axis in front of one or more calorimeters. Since the polarimeters are non-imaging, placing them in front of calorimeters, forward of the mirror focus, does not impact their performance. This design allows for simultaneous polarization and spectroscopy measurements with the same optic with the polarimeter filtering out most low-energy events. The instrument concept is shown in Figure 6, and key parameters are listed in Table 1.

The reference design's predicted instrumental sensitivity as a function of energy, assuming polarimeters in front of two focal planes in the four mirror design, is shown in Figure 7. The detection limits derived from these calculations are shown in Figure 1. We find that the reference

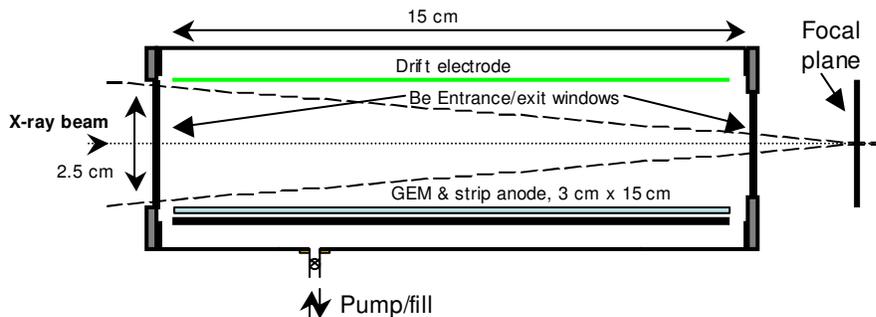

Figure 6: The TPC polarimeter concept. The design allows simultaneous polarization and spectroscopy observations.

X-ray Polarimeter for Constellation-X, Jahoda et al., response to 210S-GBG-06-001    8

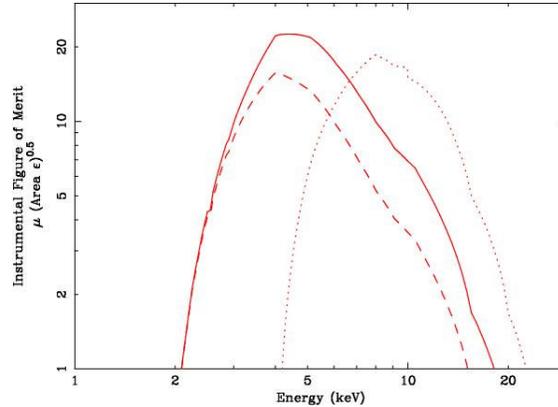

Figure 6: Instrumental sensitivity for the reference design (solid line), and for a two detector volume alternative which place 5 cm of Ne/dimethyl ether (dashes) in front of 10 cm of Ar (dots). An Ar cell would extend the polarimeter's bandpass at the cost of somewhat reduced sensitivity at low energies. Polarimeters with sensitivity at lower energies can also be constructed with pure dimethyl ether.

instrument can detect a 3% polarization in a 1 mCrab source in 40 ksec. This sensitivity opens the door to interesting observations of massive extra-galactic black holes (i.e. AGN) with modest exposures.

The energy resolution of the polarimeter is characteristic of gas proportional counters, i.e. about 20% FWHM at 6 keV. However, as polarimeters would only be in place for some of the mirror modules, the input spectrum will be simultaneously measured with excellent energy resolution.

## *5. Technology Readiness and Total Cost*

A three-year project is currently underway to advance the TPC polarimeter from its present level of development, TRL 4, to a state of flight readiness. The ROSES APRA program funds this project through FY09. Additional support is being provided by a related APRA polarimetry project and by Goddard Space Flight Center.

The major technical steps in going from the demonstration polarimeter to a flight-like detector are:

*Increasing the GEM area.* GEMs with suitable pitch and area are available from commercial sources. Development is focused on fabricating GEMs that are low-outgassing and highly breakdown-tolerant.

*Reducing the anode pitch.* A readout anode with the required 75-micron pitch was recently fabricated and is currently being integrated into a second-generation polarimeter.

*Measuring and understanding the performance of the TPC polarimeter.* A substantial effort is required to calibrate and optimize the detector. Of particular concern is the measurement and control of systematic errors. On orbit, we anticipate that tracking minimum ionizing particles will be a technique that controls many sources of systematics, for example, a change in drift velocity. A primary goal of the development project is to determine whether such monitoring techniques adequately control systematics, or if the instrument must also be rotated to achieve the required accuracy.

While it is not possible to provide a total cost over the life of the Constellation-X mission, there is high likelihood that the mission integrated cost (design, development, analysis, archiving) would be less than a few tens of millions of dollars. The instrument itself is constructed of simple components, and the eventual instrument costs will be driven by project requirements on the instruments. This instrument levies no driving requirements on the mission.